\documentclass[gmd, manuscript]{copernicus}
\nolinenumbers

\usepackage[utf8]{inputenc}
\usepackage{minted}
\usepackage{xcolor}
\definecolor{LightGray}{gray}{0.9}

\newcommand{\rulesep}{\unskip\ \vrule\ }
\newcommand{\hrulesep}{\unskip\ \hrule\ }

\begin{document}

\title{LayerCake v1.0: a framework for low- to intermediate-order automated models development}


\Author[1][Jonathan.Demaeyer@meteo.be]{Jonathan}{Demaeyer} 
\Author[2]{Ois\'in}{Hamilton}
\Author[3]{Stéphane}{Vannitsem}

\affil[1]{Meteorological and Climatological Research Service, Royal Meteorological Institute of Belgium, Belgium}
\affil[2]{Stats NZ - Tatauranga Aotearoa, Christchurch, New Zealand}
\affil[3]{School of Physical and Mathematical Sciences \& The Asian School of the Environment, Nanyang Technological University, Singapore}




\runningtitle{LayerCake v1.0}

\runningauthor{Demaeyer, Hamilton \& Vannitsem}

\received{}
\pubdiscuss{} 
\revised{}
\accepted{}
\published{}


\firstpage{1}

\maketitle

\begin{abstract}
A new model development framework called LayerCake is presented. It is written in the Python language and relies on the Sympy symbolic computation package to compute the Galerkin projection of two-dimensional spatially extended geophysical models. These projections provide their time evolution operator in the spectral space on a restricted set of modes. The resulting tendencies of the models can be evaluated either numerically (for time integration) or symbolically (for parameters sensitivity analysis and bifurcation analysis). We provide several simple examples of geophysical models to showcase the capabilities of the framework, and we hint at how more complex systems can be designed.
\end{abstract}


\introduction  

The atmosphere of the Earth constitutes a forced-dissipative dynamical system. The internal variability of the system is the result of the interaction of an immense number of processes, which span a wide range of space and time scales, forced by the radiative energy coming from the Sun. The emergent complexity of this system is so vast, it implies that evaluating its properties within a single framework is an impossible task. Reducing the number of processes that are described by models is at the center of the art of the scientific method and is at the origin of the success of atmospheric modeling and prediction.  

Since the advent of scientific computing and modern atmospheric and climate sciences, the number of models has risen extremely fast, with a course to the development of very high resolution spatio-temporal atmospheric and climate models. Besides the very high resolution models necessary for accurate quantitative description of the actual dynamics, two additional categories have always been considered as cornerstones of the understanding of the atmospheric dynamics, namely reduced- to intermediate-order models. These models are developed by limiting the description to a small set of processes interacting with each other, in order to evaluate and understand the emergence of joint dynamics~\citep{Ghil1987,Nicolis2007,Dijkstra2013}. Typically, reduced-order models have between 1 to 100 degrees of freedom, while for intermediate-order models, the number of variables covers a wider range typically from 1,000 to 100,000. These types of models have the key advantage of being easily implemented on modern computers. An additional advantage of reduced-order models is that some of their solutions can be approached analytically~\citep{Guckenheimer1983,Nicolis1995,Kuznetsov2023}. 

During the development of past reduced order models, the number of variables considered were usually small. Notable models from this period include the wind-driven transport of Henry Stommel \citep{Stommel1948} and Walter Munk \citep{Munk1950}, and the convection model of Barry Saltzman~\citep{Saltzman1962} that was used by Ed Lorenz to demonstrate the property of sensitivity to initial conditions~\citep{Lorenz1963}. For some geophysical processes, the dimensionality of the models could even be reduced to one or two variables, like the 2-box model of Stommel~\citep{Stommel1961} used to describe the dynamics of the overturning circulation in the North Atlantic, or the 0-variable Budyko-Sellers global temperature model with the albedo feedback~\citep{Budyko1969,Sellers1969}. After these first key steps in low-order modeling, there were considerable developments and multiple models of various complexities to describe atmospheric flows~\citep[e.g.][]{Pedlosky1970,Charney1980,Pedlosky1980,Cehelsky1987, Yoden1997}, ocean flows~\citep[e.g.][]{Pierini1990,Jin1990,Simonnet2003,Pierini2011} and multiscale ocean-atmosphere coupled dynamics~\citep[e.g.][]{Hasselmann1976,Frankignoul1977,Lorenz1984,Roebber1995,Vannitsem2015,Vannitsem2015GRL}, among others. 

Recently, we have developed a series of open source, reduced order multiscale models~\citep{Decruz2016,Hamilton2023,Xavier2024}, regrouped within the same code base referred to as \textsf{qgs}\footnote{For \textsf{q}uasi-\textsf{g}eostrophic \textsf{s}pectral model.}~\citep{Demaeyer2020}. These models allow the user to choose the number of modes, basis functions, and parameter values, but the governing equations for each model's component (land, ocean, atmosphere) are fixed. In the present work, we introduce a new framework in which the governing equations can also be chosen freely. It considerably extends the capabilities and flexibility by allowing the user to define their own governing equations, field projections, domain, boundary conditions and parameter values. This new framework is referred to as {\it LayerCake}.  

In the next section, this new modeling framework is introduced. Section 3 presents applications of the new framework to the barotropic regimes behaviors, to tropical dynamics, and to oversimplified planetary scale motions. The potential and extensions of the framework are discussed in the concluding section.  

\section{A new modeling framework}
LayerCake is a new python framework designed to ease the development of low- to intermediate-order geophysical models, obtained using a spectral discretization of the partial differential equations (PDEs), a procedure referred to as the Galerkin method. Due to the generally stratified nature of geophysical flows on Earth, it focuses on fields defined on two-dimensional domains. But in practice, the framework allows users to define multiple layers of PDEs, representing a coupled system of PDEs. 

The user must express each term of the equations of the PDEs in a Python script, rendering the system definition process automatable. It then relies heavily on the Sympy symbolic computation framework~\citep{meurer2017} to apply the Galerkin method on a basis, a set of user-defined spatial modes. This consists of performing symbolic or numerical integrations to obtain the inner products representing the projection of the model's PDEs onto the basis. As a result of the procedure, a set of ordinary differential equations (ODEs) are obtained, involving the projection coefficients for the model's prognostic fields. Integrating these ODEs then provide the time evolution of these coefficients, yielding the model trajectory in the (truncated) spectral space.

A particularity of LayerCake is that its output is either a Numba-accelerated Python callable~\citep{lam2015} - providing a numerical function directly - or a list of symbolic ODE tendencies that can be integrated in any of the supported languages (currently Fortran, Julia, and Python, with the possibly of adding other languages in the future). These two different outputs allow for the use of modern analysis tools, which are available in different languages~\citep{Demaeyer2025, datseris2018, doedel2007}.

\subsection{Composing partial differential equations}
LayerCake allows a user to compose and specify their system of equations piece by piece, involving basic terms $T[\psi](u,v)$ such as linear operators ($\nabla, \nabla^2, \partial_u, \partial_v, \ldots$), acting on fields $\psi$, defined over the specified spatial coordinates $u$ and $v$. Simple analytic functions $T(u,v)$ of these coordinates can also be specified. Functional compositions and products of these basic terms can also be specified as components of the equations, allowing for nonlinear behaviors to be included. Finally, some useful operators such as advection (Jacobian) and gradient of fields are also provided.

Using these building blocks provided by LayerCake, a set of two-dimensional partial differential equations (PDEs) can be defined
\begin{equation}
\label{eq:PDEs}
    \partial_t \mathcal{F}^{\mathrm LHS}_i \left[\psi_1, \ldots, \psi_N\right](u,v) = \mathcal{F}^{\mathrm{RHS}}_i \left[\psi_1, \ldots, \psi_N\right](u,v) ~ , \qquad i = 1,\ldots,N    
\end{equation}
on a particular domain specified by its $u$ and $v$ coordinates, where each functional used in the equations above - whether in the LHS or the RHS - are a sum of aforementioned terms:
\begin{equation}
 \mathcal{F}_i\left[\psi_1, \ldots, \psi_N\right](u,v) = \sum_{m=1}^{M_i} T_m^i   \left[\psi_1, \ldots, \psi_N\right](u,v)
\end{equation}
where $M_i$ is the number of terms in PDE $\mathcal{F}_i$. Above, all of the $T^i_m$ terms can depend on any of the fields $\psi_j$ of the system, allowing the PDEs to be coupled to each other.

This way, complete and closed sets of PDEs can be scripted in the LayerCake "language".

\subsection{Rationale behind LayerCake}

Once the set of PDEs and the domain have been specified, LayerCake is then able to transform this set into a system of ordinary differential equations (ODEs) with an automated Galerkin method. This method projects all the fields $\psi_j$ on given function basis $\phi_{j,k}$:
\begin{equation}
    \psi_j = \sum_{k=1}^{n_j} \psi_{j, k} \,\, \phi_{j,k} ~.
\end{equation}
The resulting discrete representation of the spatially continuous model defined by the PDEs is known as its (truncated) representation in the spectral domain. The spectral coefficients $\psi_{j,k}$ are determined by taking a specific inner product $\langle \cdot \, , \, \cdot \rangle$  defined by the user between the fields $\psi_j$ and the basis of functions $\phi_{j,k}$:
\begin{equation}
    \psi_{j,k} = \left\langle \phi_{j,k} , \psi_j \right\rangle ~.
\end{equation}
One can also apply the inner product to the equations~(\ref{eq:PDEs}) above, de facto projecting all the different terms composing them onto subspaces spanned by the basis of functions, and producing a tensor representation for each term:
\begin{equation}
    \label{eq:IPs}
        \left\langle \phi_{j,k} , T_m^i   \left[\phi_{1, k_1}, \ldots, \phi_{N, k_N}\right] \right\rangle
\end{equation}
where the rank of the tensor depends on the actual form of $T_m^i$ and its dependence on the fields $\psi_j$.

This projection procedure results in a set of ODEs for the coefficients $\psi_{j,k}$. The final goal of LayerCake is to produce numerical or symbolic representation of the tendencies of these sets of ODEs:
\begin{equation}
    \dot{\boldsymbol{u}} = \boldsymbol{f}(\boldsymbol{u})  ~.  
\end{equation}
In this context, $\boldsymbol{u}$ is a state vector consisting of the stacked spectral coefficients 
\begin{equation}
    \boldsymbol{u} = (\psi_{1,1},\ldots,\psi_{1,n_1}, \ldots, \psi_{N,1},\ldots,\psi_{N,n_N}) ~.
\end{equation}
The resulting LayerCake output is either a Python callable, or a set of strings representing the model's tendencies $\boldsymbol{f}$ (and also its Jacobian matrix $\boldsymbol{D f}$) that can be used to:
\begin{itemize}
    \item integrate the model over time
    \item perform sensitivity analysis
    \item perform bifurcation analysis
\end{itemize}
in a straightforward manner. This framework is therefore highly valuable in providing, in a systematic and reliable way, a set of dynamical equations related to a specific problem in which a limited number of equations are required.


\subsection{Known limitations \& Future developments}

There are currently some known limitations in LayerCake. The first one is that non-analytic functions are difficult to represent using the Galerkin method, and are therefore not covered by this software. This precludes for example the formulation of complex geophysical systems that involve the moisture content of the atmosphere or land. For instance, implementing a comprehensive low-order climate model proposed by Lorenz~\citep{Lorenz1984} proved difficult to the authors. Within LayerCake, these processes must therefore be parameterized in some way using some analytical functions.

Another limitation is the computational cost induced by the quadratures of the inner products~(\ref{eq:IPs}), using Sympy in symbolic mode or numerical quadratures in numerical mode. This can lead to very long computational time for models with more than one hundred variables. There is little chance to improve the symbolic part further, as it is directly related to Sympy performance, which has not been designed for these kinds of intensive computations. This limits the symbolic computation of the ODEs tendencies to low-order systems with less than hundreds variables. However, in the numerical case, these performance issues could be alleviated by using specific quadrature rules provided for instance by the user (LayerCake currently relies on the \verb|dblquad| Scipy routine~\citep{2020SciPy-NMeth} to perform the numerical quadratures). How to implement that in a streamlined fashion for the user remains to be seen. Therefore, improving some aspects related to computation performances will also be considered in the future.

About future developments, one can note that the current composition rules and objects available to compose the partial differential equations already cover a reasonable number of use cases. Nevertheless, new kind of terms could be added in the future, relying on the open-source development of the code on Github.

Additionally, leveraging LayerCake as the projection engine, the development of agents capable of translating PDEs mathematical expressions (for example in \LaTeX or Sympy formalism) to Python script could be investigated.

\section{Example of usage}

In this section, we describe a few examples of LayerCake usages, covering different domain geometries and model complexity. We start with a well-known "simple" 6-dimensional model for the extratropical atmosphere, then consider a 78-variable system of a tropical atmosphere, and finally a system describing the barotropic dynamics on the sphere. These examples provide an idea of the framework's versatility.

\subsection{A 6-dimensional barotropic model with regime behavior}

\begin{figure}
    \centering
    \includegraphics[width=1\linewidth]{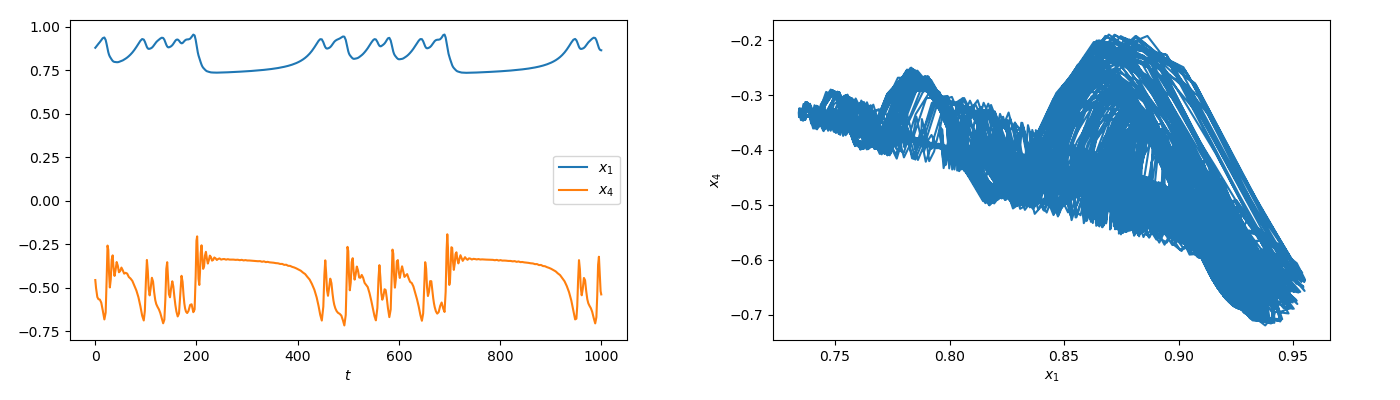}
    \hrulesep
    \includegraphics[width=1\linewidth]{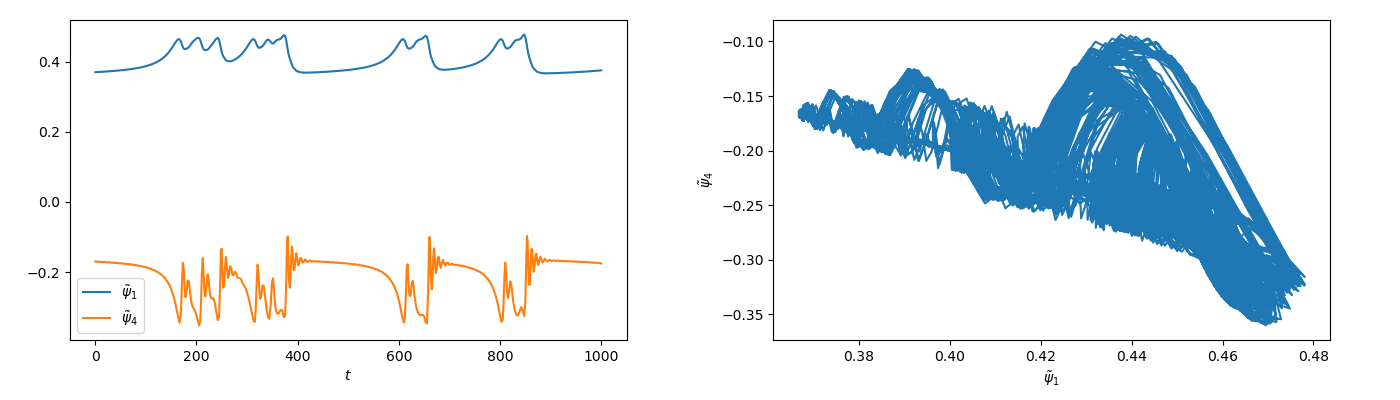}
    \caption{Top panel: Trajectory of the system~(\ref{eq:barotropic}) with the parameters found in~\cite{Crommelin2004} (in particular $b=0.5$) and $(x_1^\ast, x_4^\ast) = (0.95, -0.76095)$. Bottom panel: Trajectory of the system~(\ref{eq:CDV}) with the same parameters and $(\tilde\psi_1^\ast, \tilde\psi_4^\ast) = (0.95 \, b, -0.76095\, b)$, computed directly by LayerCake in numerical mode.}
    \label{fig:barotropic}
\end{figure}

Extratropical barotropic spectral models were proposed in the 1980s to study the interaction of planetary waves with the orography to produce stable regimes (represented by equilibrium points)~\citep{Charney1979}. Such models were later revived to propose the heteroclinic connections found in them as a plausible mechanism for regime transitions~\citep{Crommelin2004}. This model has also been used recently to study the interaction of stochastic forcing and regime dynamics~\citep{dorrington2023}. The approach offered in these works proposes to consider the barotropic equation
\begin{equation}
    \label{eq:barotropic}
    \frac{\partial}{\partial t} \nabla^2 \psi = - J(\psi, \nabla^2 \psi + f + \gamma h) - C \nabla^2 (\psi - \psi^\ast)
\end{equation}
with $f = f_0 + \beta y$ being the linearized Coriolis parameter, $J$ the usual advection operator\footnote{$J(S,G) = \partial_x S \, \partial_y G - \partial_y S \, \partial_x G$.}, and $h$ the orographic profile; and project it on the complex modes:
\begin{equation}
\label{eq:complex_modes}
\begin{aligned}
    \phi_{01}(x,y) & = \sqrt{2}\, \cos(y/b) \\
    \phi_{11}(x,y) & = \sqrt{2} \,e^{i x}\, \sin(y/b) \\
    \phi_{-11}(x,y) & = \sqrt{2} \,e^{-i x}\, \sin(y/b) \\
    \phi_{02}(x,y) & = \sqrt{2}\, \cos(2 y/b) \\
    \phi_{12}(x,y) & = \sqrt{2} \,e^{i x}\, \sin(2 y/b) \\
    \phi_{-12}(x,y) & = \sqrt{2} \,e^{-i x}\, \sin(2 y/b) \\
\end{aligned}
\end{equation}
defined on a non-dimensional beta plane $(x, y) \in [0, 2\pi] \times [0, \pi b]$.
The barotropic streamfunction $\psi$ is then given by the truncated expansion
\begin{equation}
    \psi(t, x,y) = \sum_{n=-1}^1 \sum_{m=1}^2 \psi_{nm} (t) \, \phi_{nm}(x, y)
\end{equation}
and recombining linearly these coefficients $\psi_{nm}$ into the real variables $x_i$, $i = 1, \ldots, 6$ yields finally the following system of ODEs:
\begin{equation}
    \begin{aligned}
        \dot x_1 & = \tilde \gamma_1 x_3 - C (x_1 - x_1^\ast) \\
        \dot x_2 & = - (\alpha_1 x_1 - \beta_1) x_3 - C x_2 - \delta_1 x_4 x_6 \\
        \dot x_3 & = (\alpha_1 x_1 - \beta_1) x_2 - \gamma_1 x_1 - C x_3 + \delta_1 x_4 x_5 \\
        \dot x_4 & = \tilde \gamma_2 x_6 - C (x_4 - x_4^\ast) + \varepsilon (x_2 x_6 - x_3 x_5)  \\
        \dot x_5 & = - (\alpha_2 x_1 - \beta_2) x_6 - C x_5 - \delta_2 x_4 x_3 \\
        \dot x_6 & = (\alpha_2 x_1 - \beta_2) x_5 - \gamma_2 x_4 - C x_6 + \delta_2 x_4 x_2 \\
    \end{aligned}
\end{equation}
where a certain profile for the forcing $\psi^\ast$ (purely zonal, i.e. only $x^\ast_1 \equiv \psi^\ast_{01} / b$ and $x^\ast_4 \equiv \psi^\ast_{02} /b$ are non-zero) and the orography $h$ have been assumed to be a mountain and a valley: $h(x,y) = \cos(x) \, \sin(y/b)$ (meaning $h_{11} = h_{-11} = 1/(2\sqrt{2})$). The expressions for the coefficients $\gamma_{1,2}$, $\tilde \gamma_{1,2}$, $\alpha_{1,2}$, $\beta_{1,2}$, $\delta_{1,2}$ and $\varepsilon$ can be found in~\cite{Crommelin2004}. Integrating this system with the parameters proposed in this article (in particular $b=0.5$), and with $x_1^\ast = 0.95$ and $x_4^\ast = -0.801 \, x_1^\ast$ such that it is close to a homoclinic bifurcation, which generates excursions far from one equilibrium of the system, approaching its two other equilibria without reaching them~\citep{Crommelin2004}, as seen on the top panel of Fig.~\ref{fig:barotropic}.

Using LayerCake symbolic mode, equation~(\ref{eq:barotropic}) can be coded with the following simple script:
\begin{minted}
[
% frame=lines,
framesep=2mm,
baselinestretch=1.2,
bgcolor=LightGray,
fontsize=\footnotesize,
% linenos
]
{python}
# Defining the domain
######################
b = symbols('b')
b_param = Parameter(0.5, symbol=b)
parameters = [b_param]

cs = PlanarCartesianCoordinateSystem(extent=((0, 2*pi), (0, b*pi)))


# Defining a specific basis for the model
##########################################
basis = SymbolicBasis(cs, parameters)

# coordinates
x = basis.coordinate_system.coordinates_symbol_as_list[0]
y = basis.coordinate_system.coordinates_symbol_as_list[1]

basis.functions.append(sqrt(2) * cos(y / b))
basis.functions.append(2 * cos(x) * sin(y / b))
basis.functions.append(2 * sin(x) * sin(y / b))
basis.functions.append(sqrt(2) * cos(2 * y / b))
basis.functions.append(2 * cos(x) * sin(2 * y / b))
basis.functions.append(2 * sin(x) * sin(2 * y / b))

# creating a inner product definition with an optimizer for trigonometric functions
coordinate_system=basis.coordinate_system
inner_products_definition = StandardSymbolicInnerProductDefinition(coordinate_system=coordinate_system,
                                                                   optimizer='trig', 
                                                                   kwargs={'conds': 'none'})

# Defining the fields
#######################
p = u'ψ'
psi = Field("psi", p, basis, inner_products_definition, units="[m^2][inner_products^-2]", latex=r'\psi')

# Barotropic field equation definition
#######################################

# defining the LHS as the time derivative of the vorticity
vorticity = OperatorTerm(psi, Laplacian, basis.coordinate_system)
barotropic_equation = Equation(psi, lhs_terms=vorticity)

# defining the advection term
advection_term = vorticity_advection(psi, psi, basis.coordinate_system, sign=-1)
barotropic_equation.add_rhs_terms(advection_term)

# adding an orographic term
gamma = Parameter(0.2, symbol=symbols(u'γ'), latex=r'\gamma')
hh = np.zeros(len(basis))
hh[1] = 0.5
h = ParameterField('h', u'h', hh, basis, inner_products_definition)

orographic_term = Jacobian(psi, h, basis.coordinate_system, sign=-1, prefactors=(gamma, gamma))

barotropic_equation.add_rhs_terms(orographic_term)

# adding the beta term
beta = Parameter(1.25, symbol=symbols(u'β'), latex=r'\beta')
beta_term = OperatorTerm(psi, D, x, prefactor=beta, sign=-1)

barotropic_equation.add_rhs_term(beta_term)

# adding a Newtonian cooling
C_param = Parameter(0.1, symbol=symbols('C'))
newtonian_cooling1 = OperatorTerm(psi, Laplacian, basis.coordinate_system, prefactor=C_param, sign=-1)

psi_ast_array = np.zeros(len(basis))
r = -0.801
psi_ast_array[0] = 0.95 * b_param
psi_ast_array[3] = r * psi_ast_array[0]
psi_ast = ParameterField('psi_ast', p+u'*', psi_ast_array, basis, 
                         inner_products_definition, latex=r'\psi^\ast')
newtonian_cooling2 = OperatorTerm(psi_ast, Laplacian, basis.coordinate_system, prefactor=C_param)

barotropic_equation.add_rhs_terms((newtonian_cooling1, newtonian_cooling2))

\end{minted}
partially reproduced from the corresponding example accompanying the code.
The computation of the inner products and final symbolic equations then take place, and yields a list of the said equations defining the dynamical system to be investigated.
We compare this system evolution to the original one of~\cite{Charney1979}, projecting Eq.~(\ref{eq:barotropic}) on the real modes:
\begin{equation}
\label{eq:real_modes}
\begin{aligned}
    \tilde \phi_1(x,y) & = \sqrt{2}\, \cos(y/b) \\
    \tilde \phi_2(x,y) & = 2 \, \cos(x)\, \sin(y/b) \\
    \tilde \phi_3(x,y) & = 2 \, \sin(x)\, \sin(y/b) \\
    \tilde \phi_4(x,y) & = \sqrt{2}\, \cos(2 y/b) \\
    \tilde \phi_5(x,y) & = 2 \, \cos(x)\, \sin(2 y/b) \\
    \tilde \phi_6(x,y) & = 2 \, \sin(x)\, \sin(2 y/b) \\
\end{aligned}
\end{equation}
using the inner product
\begin{equation}
    \left\langle S, G \right\rangle = \frac{1}{2\pi^2 b} \int_0^{2\pi} \int_0^{\pi b} S(x,y) \, G(x,y) \, \mathrm{d}y \, \mathrm{d}x
\end{equation}
to yield a set of equations which with $h(x,y) = h_2 \tilde\phi_2(x,y)$, $h_2 = \frac{1}{2}$ and $\psi^\ast(x,y) = \tilde\psi^\ast_1 \,\tilde\phi_1(x,y) + \tilde\psi^\ast_4 \, \tilde\phi_4(x,y)$ can be rewritten as
\begin{equation}
    \label{eq:CDV}
    \begin{aligned}
        \dot{\tilde\psi}_{1} & =  +C \tilde\psi^*_{1} - C \, \tilde\psi_{1}+ \tilde\gamma_1 \, \tilde\psi_{3}\\
        \dot{\tilde\psi}_{2} & = - C \, \tilde\psi_{2} + \beta_1\, \tilde\psi_{3} - \frac{\alpha_1}{b} \, \tilde\psi_{1}\, \tilde\psi_{3}- \frac{\delta_1}{b} \, \tilde\psi_{4}\, \tilde\psi_{6} \\
        \dot{\tilde\psi}_{3} & = - C \, \tilde\psi_{3} - \gamma_1 \, \tilde\psi_{1}- \beta_1 \, \tilde\psi_{2} + \frac{\alpha_1}{b} \, \tilde\psi_{1}\, \tilde\psi_{2}+ \frac{\delta_1}{b} \, \tilde\psi_{4}\, \tilde\psi_{5} \\
        \dot{\tilde\psi}_{4} & = +C \tilde\psi^\ast_{4} - C \, \tilde\psi_{4}+ \tilde\gamma_2 \, \tilde\psi_{6} +\frac{\varepsilon}{b} \, \left( \tilde\psi_{2}\, \tilde\psi_{6}- \tilde\psi_{3}\, \tilde\psi_{5} \right) \\
        \dot{\tilde\psi}_{5} & = - C \, \tilde\psi_{5}+ \beta_2 \, \tilde\psi_{6} - \frac{\alpha_2}{b} \, \tilde\psi_{1}\, \tilde\psi_{6} - \frac{\gamma_2}{b} \, \tilde\psi_{3}\, \tilde\psi_{4} \\
        \dot{\tilde\psi}_{6} & = - C \, \tilde\psi_{6} - \gamma_2 \, \tilde\psi_{4}- \beta_2 \, \tilde\psi_{5} +\frac{\alpha_2}{b} \, \tilde\psi_{1}\, \tilde\psi_{5}+\frac{\delta_2}{b} \, \tilde\psi_{2}\, \tilde\psi_{4} ~. \\
    \end{aligned}
\end{equation}
We see that one gets exactly the same system as in~\cite{Crommelin2004}, except for the advection terms $\tilde \psi_i \, \tilde \psi_j$, for which an extra factor $1/b$ appears in front. With a correct rescaling of the forcing $\psi^\ast$, this yields a similar dynamics, as shown on the bottom panel of Fig.~\ref{fig:barotropic}.

This case is one of the basic coding examples bundled with the LayerCake code, implemented as \\ the example \verb|barotropic_one_layer.py| in the \verb|examples| folder of the code~\citep{demaeyer2026}.

\subsection{A 78-dimensional version of the WTG Tropical Model}
\label{sec:WTG}

\begin{figure}
    \includegraphics[width=0.45\linewidth]{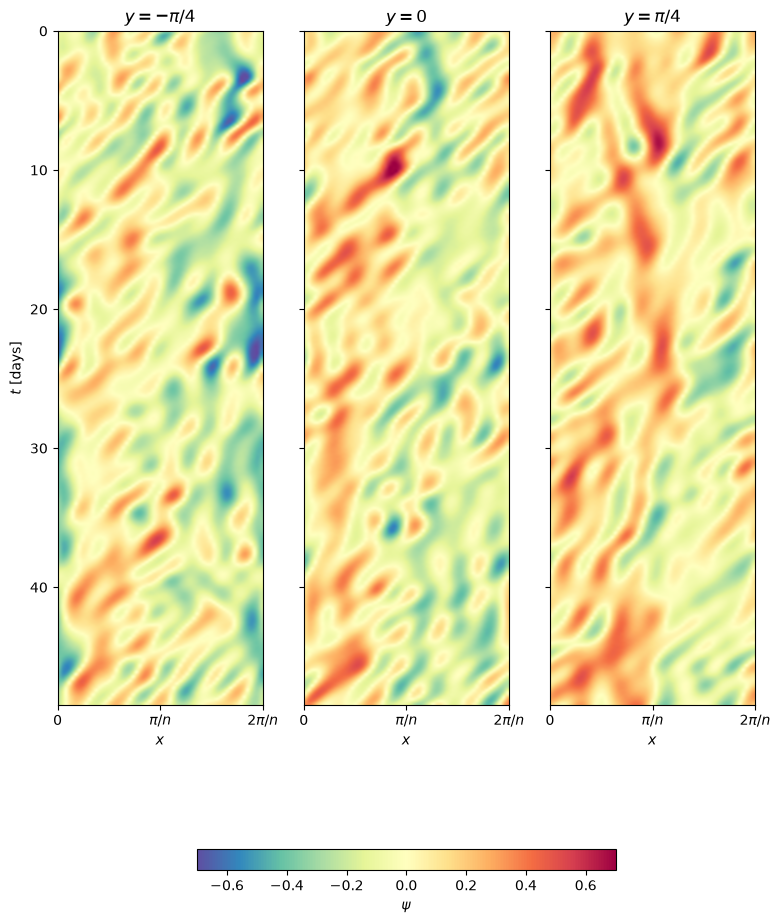}
    \rulesep
    \includegraphics[width=0.45\linewidth]{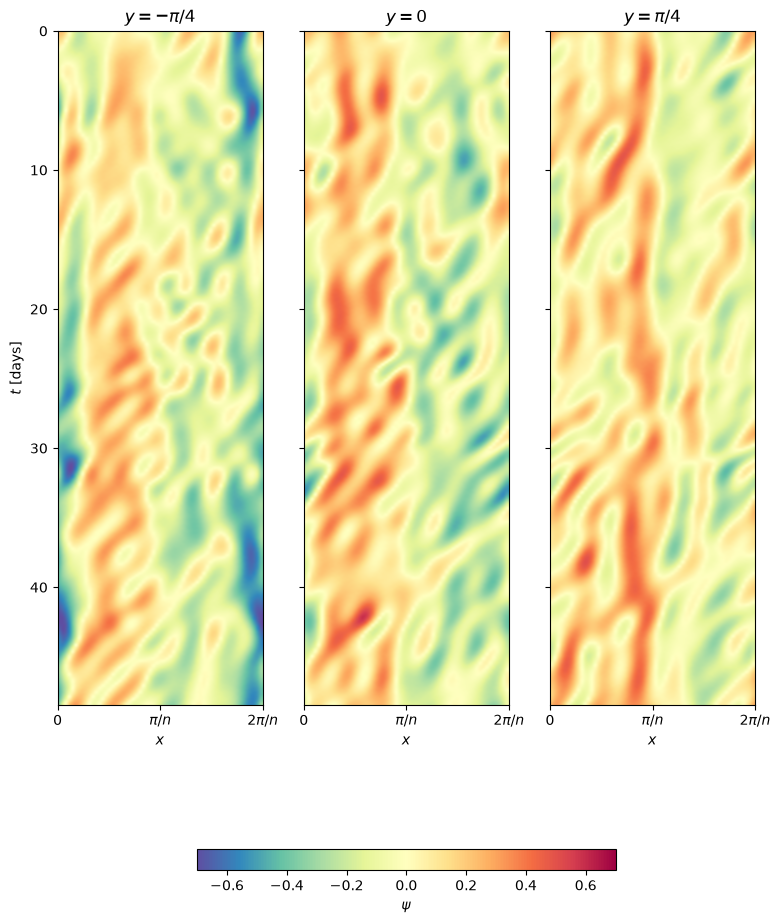}
    \caption{Left panel: Hovmöller diagram of the WTG model~(\ref{eq:nondim_WTG_PDE}), projected on 78 Fourier modes by LayerCake. Right panel: Hovmöller diagram of the WTG model~(\ref{eq:nondim_WTG_PDE}) with the addition of a turbulent diffusion term $A\nabla^2 \psi$, again projected on 78 Fourier modes.}
    \label{fig:hovmoeller}
\end{figure}

Recently, a new low-order "dry" atmospheric tropical model has been proposed by~\cite{vannitsem2026}. The model is barotropic one-layer atmosphere with a specific divergent forcing $\chi$ which results in a multiplicative forcing acting on some modes. The non-dimensional barotropic equation reads like this:
\begin{equation}
    \label{eq:nondim_WTG_PDE}
    \frac{\partial}{\partial t}  \nabla^2 \psi = - J(\psi, \nabla^2 \psi) - \beta' \frac{\partial \psi}{\partial x} - \nabla \chi \cdot \nabla (\nabla^2 \psi + f') -  (\nabla^2 \psi + f') \nabla^2 \chi - r' \nabla^2 \psi
\end{equation}
where the non-dimensional forcing $\chi$ is provided by
\begin{equation}
    \chi  = A' \cos^2 (y) \cos (n x)+ B' \sin (y) \cos (n x) \label{eq:forcing-nondim} .
\end{equation}
We direct the reader to the~\cite{vannitsem2026} reference for more details about this interesting model, in particular its stability properties. Most of the symbolic and numerical computation displayed there have been obtained by leveraging the functionalities of LayerCake. One of the latest experiments made using this model is displayed on the left panel of Fig.~\ref{fig:hovmoeller} where a Hovmöller diagram of 78 modes version of the model~(\ref{eq:nondim_WTG_PDE}) is depicted. One striking aspect of this diagram is the presence of fine, filament-like structures in the streamfunction $\psi$ plots. 
These structures were absent in the lower dimensional version of this model~\citep{vannitsem2026}. This might be partly due to the lack of dissipation at small scales as the resolution of the model is increased. In order to introduce such dissipation, we leveraged LayerCake abilities to model PDEs and added a term $A \, \nabla^4 \psi$ proportional to the square of the Laplacian to the equation~(\ref{eq:nondim_WTG_PDE}). This term models the the kinematic dissipation of energy due to turbulent diffusion at smaller scales~\citep{decruz2018, VanderAvoird2002}, and adding it only requires a few lines in the WTG model LayerCake script. The results of this addition can be seen on the right panel of Fig.~\ref{fig:hovmoeller} where the field $\psi$ appears smoothed and the remaining filaments are larger.

A final note on these two diagrams is that the waves in this model are predominently travelling westward, with some local interesting eastward propagations, a feature already noted in~\cite{vannitsem2026} and apparently surviving at higher resolution, as shown here.

\subsection{A low-order barotropic model on the sphere}

\begin{figure}
    \includegraphics[width=0.45\linewidth]{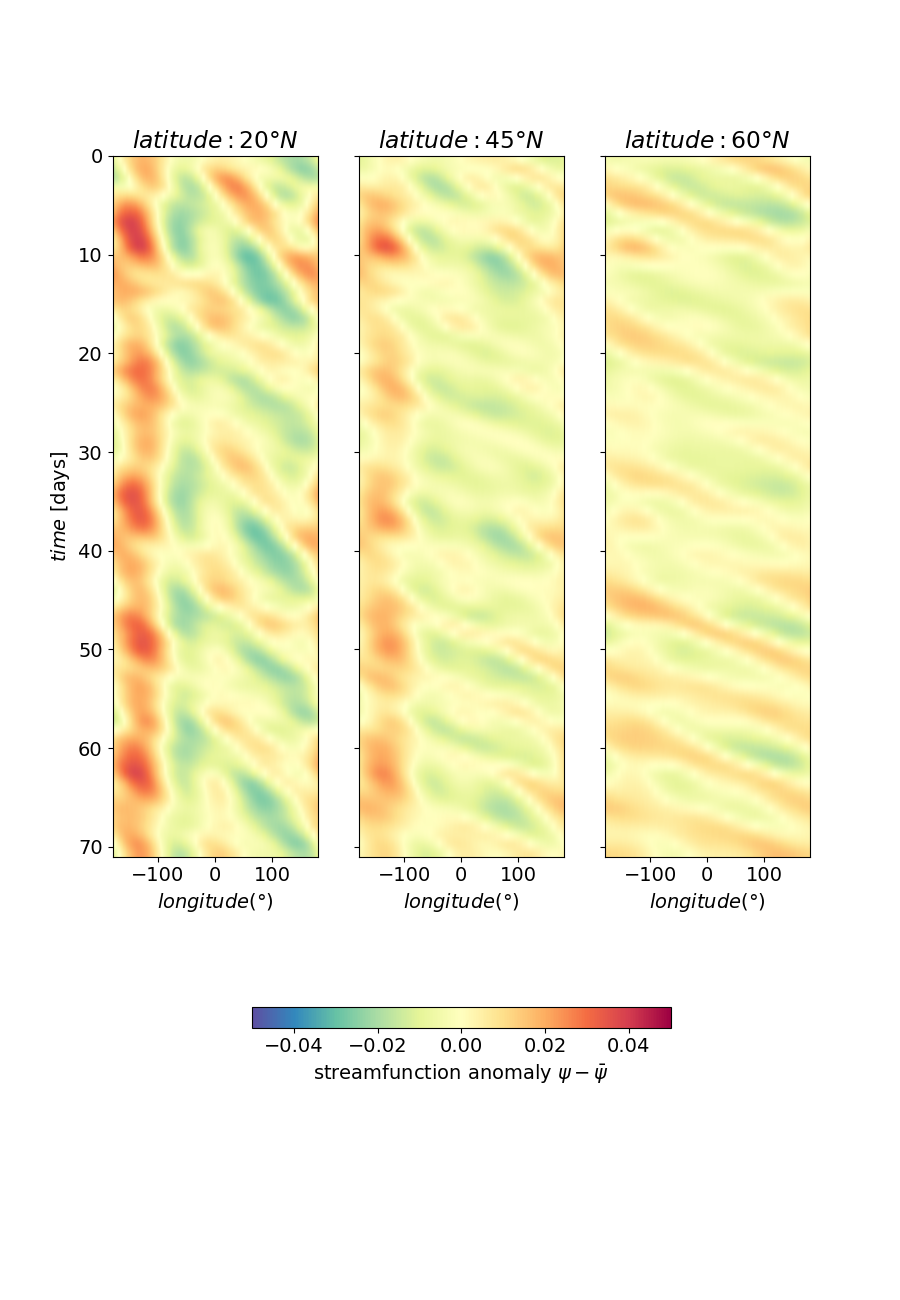}
    \rulesep
    \includegraphics[width=0.45\linewidth]{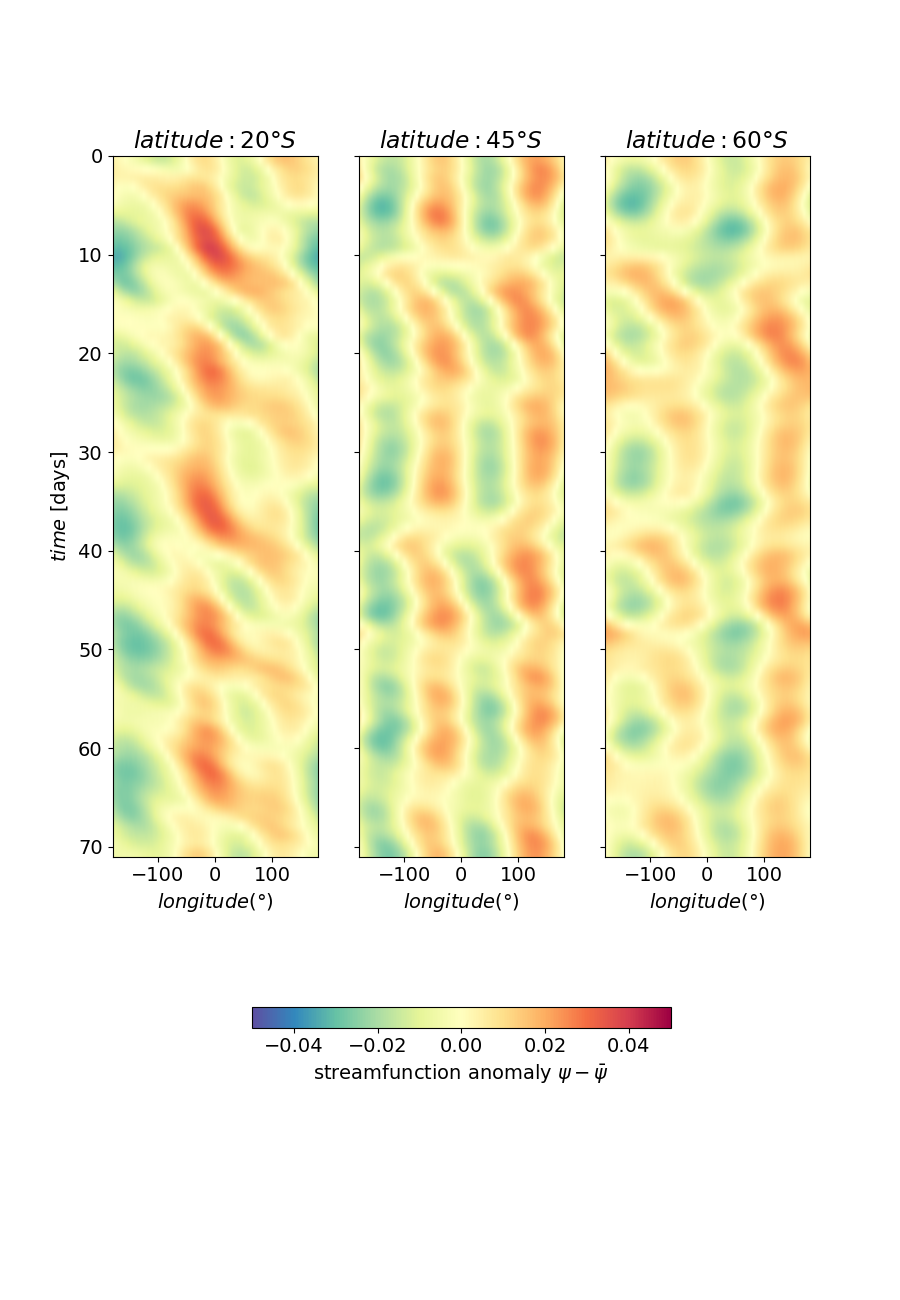}
    \caption{Left panel: Hovmöller diagram at different latitudes for the T4 truncated system~(\ref{eq:barotropic_sphere}) Northern hemisphere, showing eastward traveling waves. Right panel: Hovmöller diagram at different latitudes for the same model in the Southern hemisphere, showing stationary waves.}
    \label{fig:hovmoeller_baro_sphere}
\end{figure}

Finally, we present here an example with a geometry of the domain that is different from the planar domain of the two previous ones. We consider the following barotropic model on the sphere:

\begin{equation}
\label{eq:barotropic_sphere}
    \frac{\partial}{\partial t} \nabla^2 \psi = - J(\psi, \nabla^2 \psi + f (1 + h /H)) - C \nabla^2(\psi - \psi^\ast)
\end{equation}
which is very close to equation~(\ref{eq:barotropic}). The spherical nature of the model is then imposed by selecting spherical harmonics as the model's spectral modes. LayerCake, relying on the Sympy Python symbolic computation package, is then able to process the computation of the inner products on the sphere, showing its versatility in model development.
The LayerCake code for this simple barotropic model is quite simple and reads:
\begin{minted}
[
% frame=lines,
framesep=2mm,
baselinestretch=1.2,
bgcolor=LightGray,
fontsize=\footnotesize,
% linenos
]
{python}
# Defining the domain
######################
# Adimensional Earth sphere radius parameter
Rp = Parameter(1., symbol=Symbol("R"), units='')
parameters = [Rp]

basis = SphericalHarmonicsBasis(parameters, {'M': 4})
coordinate_system=basis.coordinate_system
inner_products_definition = StandardSymbolicInnerProductDefinition(coordinate_system=coordinate_system,
                                                                   optimizer='trig',
                                                                   kwargs={'conds': 'none'})

# coordinates
llambda = basis.coordinate_system.coordinates_symbol_as_list[0]
phi = basis.coordinate_system.coordinates_symbol_as_list[1]

# Earth rotation angular speed
omega = Parameter(7.292e-5, symbol=Symbol(u'ω'), units='[s^-1]')

# Timescale parameter
T = Parameter(1. / (2. * float(omega)), symbol=symbols('T'), units='[s]')

# Defining the fields
#######################
p = u'ψ'
psi = Field("psi", p, basis, inner_products_definition, units="[m^2][^-2]", latex=r'\psi')

# Barotropic field equation definition
#######################################

# defining the LHS as the time derivative of the vorticity
vorticity = OperatorTerm(psi, Laplacian, basis.coordinate_system)
barotropic_equation = Equation(psi, lhs_terms=vorticity)

# defining the advection term
advection_term = vorticity_advection(psi, psi, basis.coordinate_system, sign=-1)
barotropic_equation.add_rhs_terms(advection_term)

# adding an orographic term (sin theta * h)
sin_th_h = np.array([-0.001711890606304873, 0.0022332703338084825,
                     0.001014260540193253, 0.002428461633157703,
                     0.0016877961736783316, 0.0015782594096663554,
                     -0.003619193452466132, -0.006561605391195204,
                     0.0009603132306877926, -0.00197822560376025,
                     0.019943023135503446, -0.00420363112509415,
                     0.0074384102492688975, -0.014195299050184684,
                     -1.137858057111412e-05, -0.001941233058648731,
                     0.0009283207552447782, 0.0012240310990215871,
                     -0.004083320319716952, -0.007690520373983598,
                     -0.005570890927646672, -0.0014451628606521009,
                     -0.0009667440544924891, 0.002455772979125063])
oro = ParameterField('h', u'h', sin_th_h, basis, inner_products_definition)

orographic_term = Jacobian(psi, oro, basis.coordinate_system, sign=-1)

barotropic_equation.add_rhs_terms(orographic_term)

# adding the beta term
beta_term = OperatorTerm(psi, D, llambda, sign=-1)
barotropic_equation.add_rhs_term(beta_term)

# adding a Newtonian cooling
C_param = Parameter(0.00294, symbol=symbols('C'))  # corresponds roughly to a 27 days timescale
newtonian_cooling1 = OperatorTerm(psi, Laplacian, basis.coordinate_system, prefactor=C_param, sign=-1)

# somewhat exaggerated forcing (~10 times) to generate chaos in this T4 model
psi_ast_array = np.array([ 0.00144388,  0.00092015, -0.00364928, -0.01180132, -0.02419153,
       -0.0184391 , -0.02481095, -0.04143466, -0.01375496,  0.01036605,
       -0.06866103, -0.00700027,  0.00255661,  0.00643457, -0.00858923,
       -0.00973486,  0.01376331, -0.00086079, -0.02540955, -0.02291994,
       -0.02918463, -0.02469974, -0.01380329,  0.00674556])


psi_ast = ParameterField('psi_ast', p+u'*', psi_ast_array, basis, 
                         inner_products_definition, latex=r'\psi^\ast')
newtonian_cooling2 = OperatorTerm(psi_ast, Laplacian, basis.coordinate_system, prefactor=C_param)

barotropic_equation.add_rhs_terms((newtonian_cooling1, newtonian_cooling2))

\end{minted}
The full code for this model is implemented as the example \verb|barotropic_planetary_flow.py| in the examples portfolio of LayerCake~\citep{demaeyer2026} with a very low-order T4 truncation, leading to a 24-dimensional ordinary differential equation. This model resolves planetary waves 1-4 and a broad zonal-mean jet structure, but nothing smaller. An example of an integration is provided as Hovmöller diagram on Figure~\ref{fig:hovmoeller_baro_sphere}, with a somewhat arbitrary and exaggerated forcing $C\nabla^2 \psi^\ast$ and a dissipation $C$ corresponding to a 27 days timescale. This was deliberately chosen to make the nonlinear terms dominate and drive the system into a chaotic region. The diagram shows eastward propagation of waves on the Northern hemisphere midlatitudes, and stationary waves in the Southern hemisphere, probably related to forcing differences between the two (i.e. the orographic and climatological differences between the two hemispheres).

A video is also provided as supplementary material of the present article to show the evolution of streamfunction $\psi$ on the sphere of this T4 heavily truncated model~\citep{tibav:74135}.

\subsection{Additional available applications}

All the three examples presented so far consisted of only one two-dimensional partial differential equation, but as it is frequently the case in geosciences, one can discretize stratified fluid models and design system of layered partial differential equations to represent them~\citep{vallis2017}. Typical examples include baroclinic models of the midlatitude atmosphere~\citep{Charney1980,reinhold1982}, or more recently ocean-atmosphere or land-atmosphere coupled models~\citep{Vannitsem2015,Xavier2024}. 

Being able to model such layered systems is the main goal of LayerCake. In this respect, it should be noted that a two-layer baroclinic model with 10 modes (the one developed in~\cite{reinhold1982}), as well as the \textsc{maooam} ocean-atmosphere model~\citep{Decruz2016} with 18 modes have been reproduced using LayerCake, and are proposed as basic examples in the \verb|examples| folder of the code~\citep{demaeyer2026}.

\conclusions  

We have introduced here LayerCake, a new Python framework to develop systems of layered and coupled partial differential equations, which are frequently encountered in theoretical studies in geosciences. These systems are then automatically projected on sets of spatial modes by LayerCake, producing their reduced time evolution operator as sets of ordinary differential equations (typically one per spatial mode), which are easily integrable on modern computers. This work allows for the study of models of increased complexity, both in term of the number of modes and in term of the design of the system itself. Indeed, as shown in Section~\ref{sec:WTG}, one can easily add new terms to the partial differential equations to alter the model or add new processes. Here, we have presented the capabilities of LayerCake on a few well known simple atmospheric models to present the potential of LayerCake. We believe the package allows one to investigate new research questions by designing adhoc low-order models, and test the outcome as the complexity and realism is increased.

LayerCake was developed to address the limitations of previous frameworks, like \textsc{maooam} (\cite{Decruz2016}, in Fortran and Lua languages) or \textsf{qgs} (\cite{Demaeyer2020}, in Python language), where the model's number of modes and parameters can be changed by the user, but not the equations of the model itself. This forced a user, who wanted to alter these equations, to go deep into the model's code to change them, with the possibility of adding errors always present. However, the underlying Galerkin procedure generating the equations is easily reproduced by a computer, provided that it has access to a symbolic computation system. The development of Sympy~\citep{meurer2017} opened this door. The extension to even higher dimensional systems needs to be further explored by reducing the computer demand for the computation of the inner products. 

A final word must be said about the development of frontier AI models. Work on developing LayerCake started in 2024, when agentic LLM capabilities were still in their infancy. Since then, they have been shown to be able to solve open questions in research level mathematics, proving that an advanced usage of these tools for research was possible. While these tools have proven their importance in mathematical and scientific research, they are unreliable as the sole engine for long or repeated symbolic computation. In addition, there is no easy manner for a researcher to verify the resulting symbolic computation from LLMs, and repeated computations may result in different expansions of the equations, further complicating verification. Instead of asking AI to perform the Galerkin expansion, we see a potential for using AI to generate the Python scripts to run LayerCake. This would provide researchers with a fast method of building such models, while providing an easily verifiable output to inspect.

We plan to continue using and developing LayerCake for the construction of future atmospheric and climate models as done for instance in the analysis of the Weak Temperature Gradient tropical model \citep{vannitsem2026}. The scientific community is welcome to use this tool for their own research problems, and to contribute to its developments on Github.

\codeavailability{The LayerCake code is available on Github at \url{https://github.com/Climdyn/LayerCake}. Open source development will continue on this platform in the future. Additionally, it is archived on Zenodo at \url{https://doi.org/10.5281/zenodo.24166282}~\citep{demaeyer2026}. 

The code of the Tropical Weak Temperature Gradient model~\citep{vannitsem2026} used for the example in section~\ref{sec:WTG} is available on Github at \url{https://github.com/Climdyn/WTG-TM}. Additionally, it is also archived on Zenodo at \url{https://doi.org/10.5281/zenodo.20760509}~\citep{demaeyer_wtg_2026}.} 




\videosupplement{The time evolution of the streamfunction field of barotropic T4 model is provided in a video, which is available online~\citep{tibav:74135}, highlighting the modeling capabilities on the sphere provided by LayerCake} 










\authorcontribution{Inspired by O. H. contribution to the symbolic part of \textsf{qgs}~\citep{Demaeyer2020}, J.D. initiated the development of the LayerCake code. O.H. contributed to the code in some specific areas. All the authors, O.H., S.V. and J.D., have discussed and developed the examples of reduced-order model, have produced and discussed the results, and have written the manuscript.} 

\competinginterests{The authors declares no competing interest.} 


\begin{acknowledgements}
S. Vannitsem is partly supported by the Ministry of Education, Singapore, under Academic Research Fund (AcRF) Tier 1 Grant RG15/26.
\end{acknowledgements}







\bibliographystyle{copernicus}
\bibliography{references.bib}

\end{document}